# The EAGLE instrument for the E-ELT: developments since delivery of Phase A

Simon L. Morris[*a], Jean-Gabriel Cuby[b], Marc Dubbeldam[a], Christopher Evans[c], Thierry Fusco[d], Pascal Jagourel[e], Richard Myers[a], Phil Parr-Burman[c], Gerard Rousset[f], Hermine Schnetler[c]

[a] Department of Physics, Durham University, South Road, Durham, DH1 3LE, UK, [b]Aix Marseille Université, CNRS, LAM UMR 7326, 13388, Marseille, France [c] UK Astronomy Technology Centre, Royal Observatory Edinburgh, Blackford Hill, Edinburgh, EH9 3HJ, UK, [d]ONERA, France, [e] GEPI, Observatoire de Paris, 5 place Jules Janssen, 92195 Meudon Cedex, France, [f] LESIA, Observatoire de Paris, 5 place Jules Janssen, 92195 Meudon Cedex, France

## ABSTRACT

The EAGLE instrument is a Multi-Object Adaptive Optics (MOAO) fed, multiple Integral Field Spectrograph (IFS), working in the Near Infra-Red (NIR), on the European Extremely Large Telescope (E-ELT). A Phase A design study was delivered to the European Southern Observatory (ESO) leading to a successful review in October 2009. Since that time there have been a number of developments, which we summarize here. Some of these developments are also described in more detail in other submissions at this meeting.

The science case for the instrument, while broad, highlighted in particular: understanding the stellar populations of galaxies in the nearby universe, the observation of the evolution of galaxies during the period of rapid stellar build-up between redshifts of 2-5, and the search for 'first light' in the universe at redshifts beyond 7. In the last 2 years substantial progress has been made in these areas, and we have updated our science case to show that EAGLE is still an essential facility for the E-ELT. This in turn allowed us to revisit the science requirements for the instrument, confirming most of the original decisions, but with one modification.

The original location considered for the instrument (a gravity invariant focal station) is no longer in the E-ELT Construction Proposal, and so we have performed some preliminary analyses to show that the instrument can be simply adapted to work at the E-ELT Nasmyth platform.

Since the delivery of the Phase A documentation, MOAO has been demonstrated on-sky by the CANARY experiment at the William Herschel Telescope.

## 1. INTRODUCTION

The EAGLE instrument for the E-ELT is meant to be a workhorse multi-object spectrograph operating in the optimal wavelength range for the telescope and its adaptive optics, and which is also well matched to the primary science drivers for the telescope as a whole (Cuby et al. 2010). It will be the only instrument that actually takes full advantage of the E-ELT's 5 mirror design, delivering data with high levels of AO correction across the full diameter of the telescope focal plane. For those familiar with instrumentation on the VLT, it will functionally provide a cross between the highly successful SINFONI AO-fed IFS spectrograph, and the (soon to arrive) KMOS IFS multiplex capability.

## 2. EAGLE SCIENCE CASE UPDATES

When delivered in October 2009, the EAGLE science case looked in detail at three main cases, all taken from the E-ELT design reference mission: Stellar populations in nearby galaxies, the assembly of galaxies during the period of peak star formation density (redshifts 2<z<7), and the properties of 'first-light' galaxies with redshifts z>7 (Evans et al. 2010). We revisit these cases below, noting any new results which might be relevant from the last 3 years.

*simon.morris@durham.ac.uk; phone 0044 191 334 3611;

## 2.1 Stellar Populations

One of the key components of the EAGLE Science Case was spectroscopy of extra-galactic resolved stellar populations. An example 'Large Programme' developed during the Phase A study was for spectroscopy of evolved stars in galaxies in the Sculptor Group (see Evans et al. 2010). The objective in this case is to determine the star-formation and mass-assembly histories of the target galaxies via spectroscopy of large numbers (100s) of individual stars, thus enabling tests of galaxy evolution models.

The key diagnostic feature for such a programme was the calcium triplet (CaT, with absorption lines at 850, 854, and 866 nm). The CaT has become a common method to obtain estimates of metallicities and stellar kinematics of evolved populations (e.g. Tolstoy et al. 2001; Koch et al. 2007), and was the prime driver for the inclusion of the shorter-wavelength region (0.80-0.90μm) in the Phase A design of EAGLE (see Evans et al. 2010). Observations of the CaT also motivated the inclusion of a second resolving power in the design of the EAGLE spectrographs, with $R\sim4000$ being too low for sufficiently precise estimates of abundances and radial velocities. The baseline value adopted for a secondary resolving power was $R\sim10,000$ – a trade-off between sensitivity versus velocity precision, plus the potential benefit of greater resolving powers to other supplementary cases, e.g., studies of stellar clusters and Galactic stars. However, if we focus solely on deriving metallicities from the CaT, a resolving power $\geq 5000$ is sufficient to recover accurate metallicity estimates (see Battaglia et al. 2008, Starkenburg et al. 2010) given adequate S/N ($\geq 20$). Thus, when revisiting the spectrograph design, prompted by some feedback from the Phase A review, we adopted an intermediate resolving power of $R\sim8000$.

While the CaT is envisaged as the 'workhorse' method for stellar population studies, since the Phase A study we have also investigated the potential of diagnostic features in the J-band, in which the AO performance will be better than at shorter wavelengths. To obtain metallicities of extra-galactic red supergiants (RSGs), Davies et al. (2010) suggested using the 1.15-1.22μm region (which includes absorption lines from Mg, Si, Ti, and Fe) at spectral resolving powers of a few thousand. Employing some of the simulation tools developed for the EAGLE study, Evans et al. (2011) investigated the potential of these J-band diagnostics to obtain metallicities of both RSGs and evolved red giant branch (RGB) stars in external galaxies with the E-ELT. They concluded that a continuum signal-to-noise in excess of 50 (per two-pixel resolution element) was required to recover the simulated input metallicity to within 0.1 dex, sufficient for many extra-galactic applications. Evans et al. concluded that direct estimates of stellar metallicities (over the range $-1 < [Fe/H] < 0$) should be possible with EAGLE on the E-ELT out to distances of ~5 Mpc for stars near the tip of the Red Giant Branch, and out to tens of Mpc for Red Super-Giants. From comparisons with the results of CaT simulations, when the improved image quality from AO and intrinsic red colours of the stars are taken into account, the J-band diagnostics appear competitive in terms of sensitivity for a given target. Further work in this area is ongoing with X-Shooter on the VLT (Davies et al. in prep).

## 2.2 Galaxy Assembly (2<z<7)

In a recent and comprehensive review, Shapley (2011) listed some of the most pressing and open questions about the physical properties of galaxies at redshifts 2<z<4. This is the redshift range which is now being surveyed with (currently single object) integral field spectrographs. These objects are initially found using a number of photometric colour techniques, and are commonly referred to as Lyman Break Galaxies (LBGs), Distant Red Galaxies (DRGs), BzK galaxies (referring directly to the photometric bands used) and Lyman alpha emitters (LAE, normally found using a narrow band filter chosen to lie in a clean part of the atmospheric emission spectrum). Alternative methods for finding galaxies over this redshift range include x-ray, sub-mm and radio selection techniques. Essentially all of these selection techniques then require follow-up spectroscopy to confirm the redshift, and increasingly now lead to spatially resolved (and preferably AO corrected) spectroscopy to derive physical properties such as dynamics, star formation densities and metallicities.

In practice, metallicity measurements for high redshift galaxies have proven very challenging. Emission line diagnostics such as the N2 indicator (log([NII] λ6584/Hα λ6563) or R23 (log(([OIII] λ5007+[OII] λ3727)/Hβ λ4861) can only be obtained for a limited set of redshift ranges (see Figure 1). Stellar absorption features from the rest frame optical (including those mentioned in section 2.1 above), and also the rest frame UV such as FeIII λ1978 or the CIV λ1550 P-Cygni feature require much higher S/N to detect. Clearly large samples of such measurements will only be possible with E-ELT class telescopes with a multi-object capability.

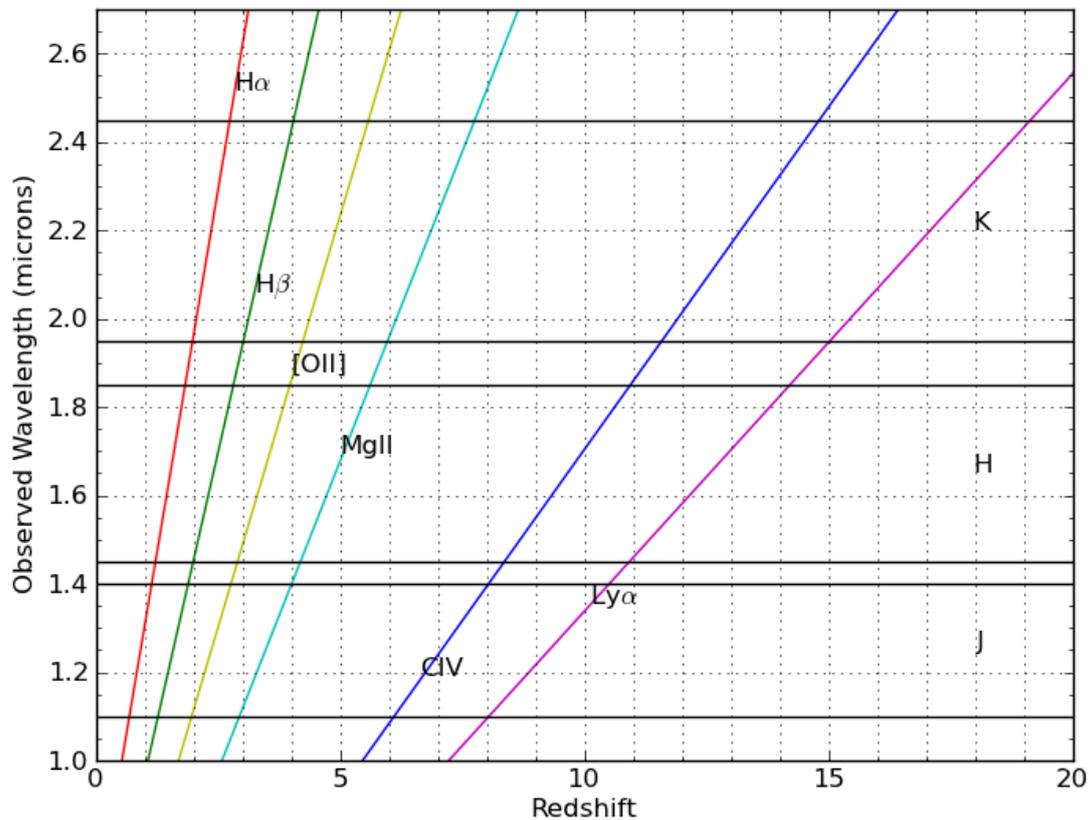

Figure 1. Redshift ranges of interest, given the atmospheric transmission bands, and the rest frame wavelengths of diagnostic emission and absorption features.

Observing one object at a time, with exposures typically taking most of a night, some groups have built up IFU samples of 10-60 objects (Förster Schreiber et al. 2009, Law et al. 2009, Mancini et al. 2011), taken from the bright end of the luminosity function, at redshifts below 2.6. A high velocity dispersion has been seen in spatially resolved spectroscopy, which is of considerable interest, but does also make classification of galaxies into (possibly naïve) classes such as 'rotating' or 'merging' difficult.

Observing strongly lensed galaxies is now well established as a way to predict what might be seen in future (bigger) samples of unlensed galaxies with larger telescopes. At present, this is the only way to obtain spatially resolved information (even from luminous galaxies) at redshifts near 5. For example, Swinbank et al. (2009) show that for one lensed, redshift 4.9, galaxy, the star formation rate density within its HII regions is approximately two orders of magnitude greater than those observed in local spiral/starburst galaxies, while still being consistent with the most massive HII regions in the local Universe, such as 30 Doradus. This may help to explain the high velocity dispersions seen in the galaxies at lower redshifts mentioned above.

As summarized by Shapley (2011), the key requirements for further understanding are:
- Larger statistical samples, by at least an order of magnitude
- Extending studies towards fainter luminosities
- Looking for evidence of accretion and outflows
- Looking into the effects of environment at high redshift

Based on the above (and other) recent references, it seems clear that, despite exciting new results on galaxies at redshifts around 2-3, the need to build up larger samples, to push down the luminosity function, and to go to higher redshifts, all mean that the need for an EAGLE like instrument on the E-ELT to study galaxy assembly is still very strong. Revisiting the science requirements, we note that the multiplex proposed for EAGLE has always been constrained by cost, and although in principle a value higher than 20 could be usable for this case, the speed gain from a multiplex of 20 is thought to still be the best match between survey speed and affordability. New information about clumpy star formation in high redshift galaxies confirms our original goal of the highest practical AO performance, and hence that the original EAGLE science requirements are still correct for this science case.

## 2.3 First Light (z>7), or galaxies in the 1$^{st}$ Gyr of the Universe.

The first substantial samples of candidate galaxies with redshifts above 7 were just appearing at the time of the delivery of the EAGLE Phase A report. Since then, huge numbers of papers have been published in the science area, Dunlop (2012) provides a very up to date and complete review. Selection of these objects uses the same approaches as described in the previous section, but at wavelengths shifted further into the Infra-red. Spectroscopic confirmation of photometrically selected targets is rare, and becomes essentially impossible, with current facilities, at the highest redshifts probed. A key issue here is the possible trapping of Lyman-α by a neutral Inter-galactic medium at z>7, potentially making this primary diagnostic feature invisible.

As an example of the new information now available, Figure 2 shows power-law fits to the mean diameter containing half of the light of LBG from Oesch et al. (2010) for two different LBG luminosity ranges as a function of redshift.

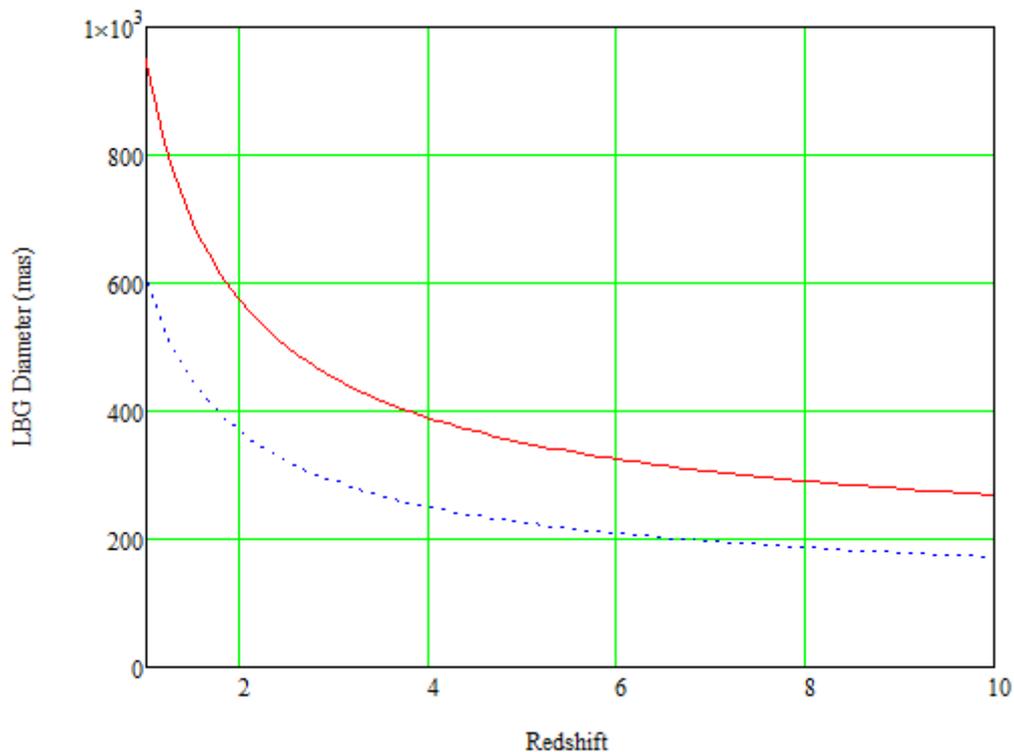

Figure 2. Mean Lyman Break Galaxy half-light **diameter** in milli-arcseconds as a function of redshift (Oesch et al. 2010). The red solid line is for LBG with (0.3-1) $L^{*}_{(z=3)}$ (where $L^{*}_{(z=3)}$ is the characteristic luminosity of LBGs from Schecter function fits to the luminosity function at z=3), while the blue dashed line is for (0.12-0.3) $L^{*}_{(z=3)}$.

The field of view of an individual Integral Field Unit (IFU) is always a trade-off between: the desire to include the whole of a given target (ideally along with some sky regions to allow background subtraction), the spatial sampling required, and the practical cost of the detector real-estate needed. As shown by Figure 2, our original selection of 1.65 x 1.65 arc-seconds still seems appropriate. Updated simulations of the exposure times needed are in progress as part of an ESO sponsored delta-Phase A study which is about to start. There will use the new information about the object structure, and also updated performance information about the E-ELT MOAO performance, based on end-to-end modeling coupled with the new data from CANARY (see below). However, it seems likely that our original selection of 37.5 milli-arcseconds spatial sampling, with an AO performance specification based on the encircled energy entering a 2 x 2 spatial sample aperture, is likely to still be appropriate.

Samples of z>7 LBG candidates selected using HST data sensitive to $m_{AB}$=29, over areas of a few 10s of square arcminutes, now contain ~70 objects at z~7, ~50 objects at z~8 and one possible detection at z~10. LAE selection at z>7 has so far been largely unsuccessful (Clement et al. 2012). Dunlop (2012) flags the following as some of the key unanswered questions about the highest redshift galaxies:

- Measurement of the faint-end slope of the high redshift luminosity function (allowing determination of the origin of the ionizing radiation causing the re-ionization of the universe.)

- Determination of the evolution of the Lyman-α luminosity function at z>7, giving information about the final stages of re-ionisation.

- Estimation of stellar initial mass function, star formation history, metallicity and dust content of z>7 galaxies to allow determination of accurate total current stellar masses, and then also specific star formation rates (i.e. star formation rates per unit stellar mass already in place)

As noted by Dunlop 2012, one of the key observational goals is therefore to obtain complete spectroscopic follow-up of LBGs, over a wide range of UV luminosity and redshift. Even if, for the fainter objects, one has to co-add all of the flux in order to push up the S/N, spatially resolved observation is probably still needed to allow optimal co-addition of the flux along with accurate background subtraction, for these extremely faint objects. For the brighter objects, spatially resolved observations could be used to begin to untangle the physics of star formation at these very early times.

Based on the above, the multiplex chosen seems well matched to the observed source densities on the sky, and the IFU properties (FoV, spatial sampling, AO performance) also still seem appropriate.

## 3. LATEST EAGLE SCIENCE REQUIREMENTS

Pleasingly, the only substantive change to the science requirements for EAGLE arising from the Phase A reviews, and also the revisit of the science cases described above, is a change to the spectral resolving power. Our Phase A design accommodated resolving powers of 4000 and 10,000. It now seems both scientifically acceptable, and also technically possible, to have spectrographs with a single resolving power of R~8000. With 4000 detector pixels in the wavelength direction, this will allow observations of a single IR band (J, H or K) with a single setting. This makes EAGLE into essentially a 'single mode' instrument, with only one spatial and one spectral sampling.

Table 1: Current EAGLE Science Requirements.

| Parameter | Value |
| --- | --- |
| Patrol Field | Equivalent to 7 arcminutes diameter |
| Science Field of each target (instantaneous science field of view) | 1.65 x 1.65 arc seconds<br>~ (5.2 x 5.2 mm in the focal plane) |
| Multiplex (number of science channels) | 20 IFS |
| Spatial pixel scale | 37.5 mas |
| Spatial resolution | Ensquared Energy ≥ 30% in 2 x 2 spatial pixels in the H Band |

| Spectral resolving power (R) | 8000 |
| Wavelength Coverage | 0.8 – 2.45 μm |
| Clustering/Tiling | Distributed/clustered targets |

## 4. MOVING EAGLE FROM THE GRAVITY INVARIANT FOCAL STATION TO THE NASMYTH

See Schnetler et al. 8446-298, Cochran et al. 8450-43 and Basden et al. 8447-100 from this SPIE meeting for more information about this.

As part of the reduction in costs for the E-ELT, the Gravity Invariant Focal Station (GIFS) from the original design has been removed. For its Phase A study, EAGLE was asked to use this focal station to provide feedback on its utility. With the removal of this focal station to save costs, the EAGLE team is now working to demonstrate that it will be possible to deploy EAGLE at one of the Nasmyth foci. As a first step in this, we have simply taken the Phase A EAGLE design, and looked at the new support system needed to hold it horizontally, and also looked at the resulting flexure. No optimization of the end-to-end design has been performed yet. We will be performing this analysis as part of a delta Phase A, starting summer 2012, with a completion expected in June 2013.

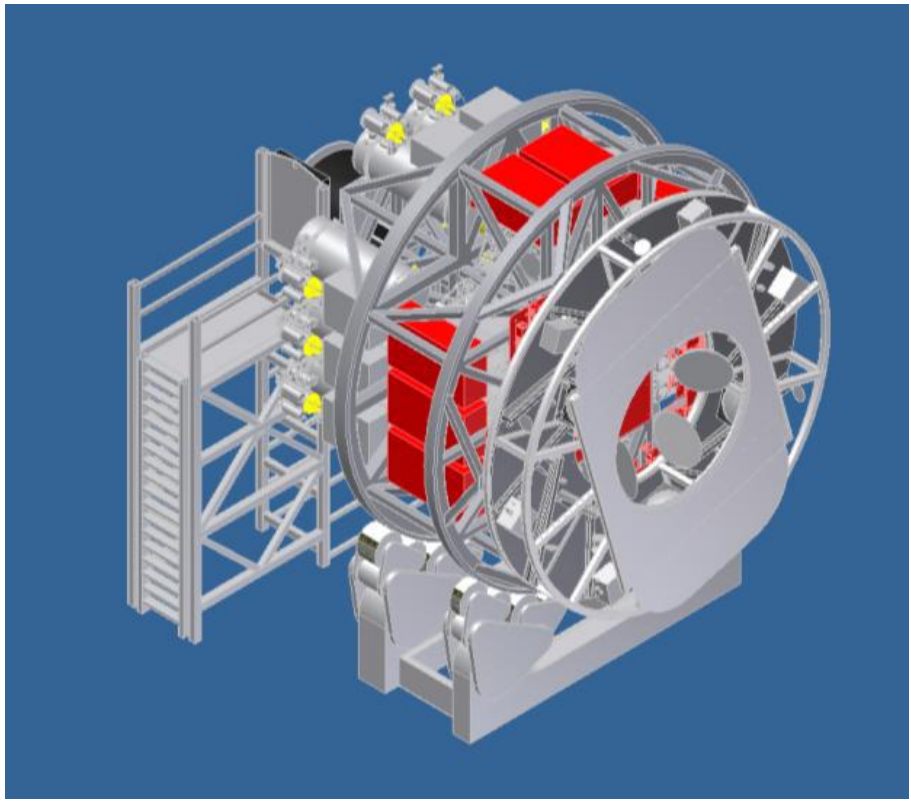

Figure 3. New Mechanical Mounting for EAGLE at the E-ELT Nasmyth Focus.

The new space frame design (see Figure 3) accommodates out of plane bending moments induced by a varying gravity vector. The design also features a high structural efficiency (stiffness-to-mass ratio). Equipment bays are provided to accommodate the various sub-systems. The frame also facilitates easy access and modularity. The frame is built up from rectangular hollow cross sections. The use of structural steel reduces the cost and eases the manufacture of the frame (e.g. welding). Instrument rotation is facilitated by two 5.5 m diameter bearings. The bearings are supported by two sets of adjustable bearing support roller blocks, containing two rollers each. One roller is driven through a gearbox to provide instrument rotation (friction drive). Synchronization with telescope is achieved by means of a signal from the Telescope Control System.

A Finite Element Model was generated and analyzed using the Autodesk Simulation Mechanical software package. Beam elements were used to model the structure; with various sub-systems represented by lumped masses. Roller supports were implemented by adding "infinitely" stiff truss elements which constrain appropriate nodes in the radial direction only; while additional restraints constrain the longitudinal and rotational solid body motion. Various load cases were analyzed for various rotation angles of the instrument.

The deformations seen are dominated by "rigid body" motion of the instrument, most notably sag in the y-direction and rotation about the x-axis. Residual movements of various subsystems relative to each other (e.g. "alignment errors") are an order of magnitude smaller, with displacements of ~0.1 mm and rotations of ~0.05 mrad. Rigid body motion can be reduced by improving the stiffness of the main bearings and supports. A more radial optical lay-out (which is possible due to the different space envelope available at Nasmyth) would facilitate a more compact instrument design and move the Centre of Gravity (CofG) closer towards telescope interface, thus reducing the load on the rear bearing and deformations of Instrument Support Structure. The possible new simplified Instrument Support Structure design, with reduced mass, would also move the CofG towards the telescope interface and reduce the load on the rear bearing and the ISS equipment bay.

In conclusion, the initial Finite Element Analysis results demonstrate that flexure is not likely to compromise instrument performance. Options have been identified to re-configure the optical design with a view to better exploiting the new space envelope;- this will likely reduce the instrument mass and facilitate a more efficient structural design with improved flexure characteristics.

## 5. RECENT HIGHLIGHTS FROM THE MOAO DEMONSTRATOR: CANARY

See Morris et al. 8447-20, Sivo et al. 8447-106, Gratadour et al. 8447-133 and Martin et al. 8447-218 at this SPIE meeting for more information about this.

The Canary MOAO demonstrator is designed to create a single MOAO channel (resembling EAGLE as closely as possible) using the 4.2m William Herschel Telescope. It is effectively a 1/10th scale model of E-ELT using 4 x 10km Rayleigh Laser Guide Stars (LGS) to emulate the E-ELT >80km Na LGS. The primary requirements are:

- Perform NGS, then LGS based tomographic Wave-front Sensing
- Perform open-loop AO correction on-sky
- Develop calibration and alignment techniques
- Fully characterize system and subsystem performance
- No imposed requirement to deliver astronomical science (i.e. no strong push for the highest possible Strehl Ratio – but including a goal to quantify the contributions of all the different error sources, and obtain a match with end-to-end modeling.)

There are currently four planned 'phases' to the project:

- Phase A: Natural Guide Star Multi-Object Adaptive Optics (essentially complete)
- Phase B: Low order Laser Guide Star Multi-Object Adaptive Optics (in progress)
- Phase C1: Ground Layer AO, Layer-Oriented Tomographic AO, Single Conjugate AO (2013/2014)

- Phase C2: Full EAGLE simulation, with an open loop high order Deformable Mirror. (2014+)

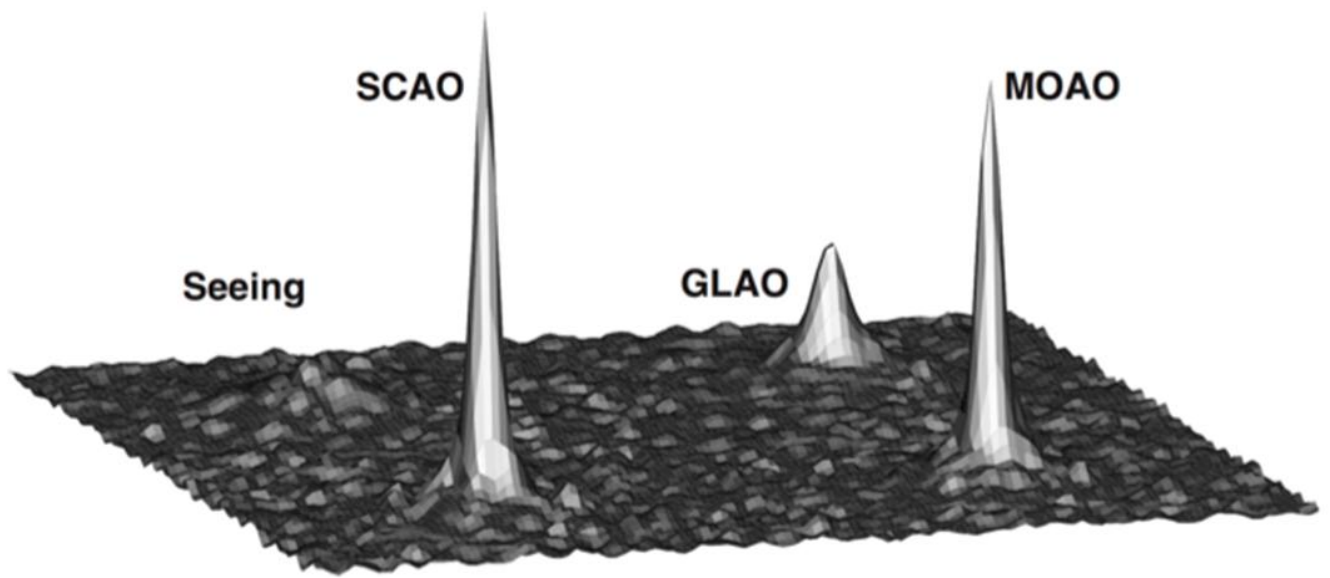

Figure 4. Point Spread Functions measured during the CANARY Phase A (see Gendron et al. 2011).

Some of the Phase A results have been published in Gendron et al. (2012). Figure 4 shows the point spread functions delivered with NGS MOAO. An analysis of the contributions to the error budget show a close match with those expected from previous modeling (see Vidal et al. 2011).

We also note that there has been work by both Phase-A teams looking at the possibility of merging the EAGLE and OPTIMOS-EVE instruments into a single workhorse E-ELT MOS. Some of this work is described in Evans et al. 8446-291 and Basden et al. 8447-200 at this SPIE meeting.

## 6. CONCLUSIONS

The EAGLE Science Case is still valid, indeed stronger than before, with a better understood need to push to lower luminosities, and higher redshifts, with larger samples, than is possible with current facilities. The spatial sampling, IFU field of view and level of AO correction still seems well matched to the primary targets. The move of EAGLE to the E-ELT Nasmyth Focus is possible. MOAO has been demonstrated on sky using natural guide stars, with further risk reduction including laser guide stars and a full demonstration of the EAGLE AO path planned.